 \documentclass[10pt,preprint]{aastex}  
 \usepackage{natbib}
\def\etal{{\it et al.}}

\def\ie{{\it i.e.,~}}

\def\arcsec{\hbox{$^{\prime\prime}$}}

\def\gapprox{\lower.4ex\hbox{$\;\buildrel >\over{\scriptstyle\sim}\;$}}
\def\lapprox{\lower.4ex\hbox{$\;\buildrel <\over{\scriptstyle\sim}\;$}}

\shortauthors{ASCHWANDEN 2012}
\shorttitle{NLFFF MODELING WITHOUT STEREOSCOPY}

\begin{document}

\title{		Nonlinear Force-Free Magnetic Field Fitting to
		Coronal Loops with and without Stereoscopy
		\footnote{Manuscript version, 2012-Nov-15}}

\author{        Markus J. Aschwanden }

\affil{         Lockheed Martin Advanced Technology Center,
                Solar \& Astrophysics Laboratory,
                Org. ADBS, Bldg.252,
                3251 Hanover St.,
                Palo Alto, CA 94304, USA;
                e-mail: aschwanden@lmsal.com}

\begin{abstract}
We developed a new nonlinear force-free magnetic field (NLFFF) 
forward-fitting algorithm based on an analytical approximation
of force-free and divergence-free NLFFF solutions, which requires
as input a line-of-sight magnetogram and traced 2D loop coordinates
of coronal loops only, in contrast to stereoscopically triangulated
3D loop coordinates used in previous studies. Test results of
simulated magnetic configurations and from
four active regions observed with STEREO demonstrate that NLFFF
solutions can be fitted with equal accuracy with or without stereoscopy,
which relinquishes the necessity of STEREO data for magnetic modeling
of active regions (on the solar disk). The 2D loop tracing method
achieves a 2D misalignment of $\mu_2=2.7^\circ\pm 1.3^\circ$ between
the model field lines and observed loops, and an accuracy of
$\approx 1.0\%$ for the magnetic energy or free magnetic energy ratio.
The three times higher spatial resolution of TRACE or SDO/AIA (compared
with STEREO) yields also a proportionally smaller misalignment angle 
between model fit and observations. Visual/manual loop tracings are 
found to produce more accurate magnetic model fits than automated 
tracing algorithms.  The computation time of the new forward-fitting 
code amounts to a few minutes per active region.
\end{abstract}


\section{          	    INTRODUCTION  			}

The success or failure of magnetic field modeling of the solar corona depends
on both the choice of the theoretical model, as well as on the choice of the 
used data sets. The simplest method is potential field modeling, which 
requires only a line-of-sight magnetogram $B_z(x,y,z_{phot})$, but only few 
solar active regions match a potential field model. Also linear force-free
field (LFFF) models are generally considered as unrealistic, where 
a single constant value for the force-free parameter $\alpha$ represents the 
multi-current system of an entire active region in the solar corona. 
The state-of-the-art is the nonlinear force-free field (NLFFF) model, which 
can accomodate for an arbitrary configuration of current systems, described 
by a spatially varying $\alpha({\bf x})$ parameter distribution 
in an active region. The next strategic decision is the choice of data sets 
to constrain the theoretical model. NLFFF models generally require vector 
magnetograph data, ${\bf B}(x,y,z_{phot})=[B_x(x,y,z_{phot}), 
B_y(x,y,z_{phot}), B_z(x,y,z_{phot})]$, which are used as a lower boundary 
constraint and are extrapolated into the corona.
However, a fundamental problem that has been identified is that the zone in the
photosphere and lower chromosphere is not force-free (Metcalf \etal, 1995), 
which spoils the extrapolation into coronal heights and leads to a substantial 
mismatch between the extrapolated field lines and observed coronal loops, 
typically amounting to a 3D misalignment angle of $\mu \approx 24^\circ -
44^\circ$ (DeRosa \etal, 2009; Sandman \etal, 2009). Obviously, this 
fundamental problem can only be circumvented by using additional constraints 
from coronal data, since the corona above the transition region is generally 
force-free, except when the plasma-$\beta$ parameter is larger than unity 
(e.g., in filaments) or when the equilibrium brakes down (e.g., during 
filament eruptions, flares, and coronal mass ejections). 

The problem amounts now how to implement coronal data into a theoretical 
magnetic field model, such as coronal loops, which are believed to be reliable 
tracers of the coronal magnetic field. A feasible approach is the method of
stereoscopic triangulation, which can be applied by using the solar rotation or
stereoscopic observations from dual spacecraft, as provided by the
STEREO mission (for reviews of solar stereoscopy see, e.g., Inhester 2006;
Wiegelmann \etal, 2009; Aschwanden 2011). Stereoscopically triangulated coronal
loop coordinates $[x(s), y(s), z(s)]$ (as a function of the curvilinear 
abscissa $s$) have been used to constrain: (i) potential field models in terms 
of buried unipolar magnetic charges (Aschwanden and Sandman
2010) or buried dipoles (Sandman and Aschwanden 2011), (ii) linear force-free
fields (Feng \etal, 2007; Inhester \etal, 2008; Conlon and Gallagher 2010),
and (iii) nonlinear force-free fields (Aschwanden \etal, 2012a). The proof 
of concept to fit NLFFF codes to prescribed field lines was also demonstrated 
with artificial (non-solar) loop data, fitting either 3D field line coordinates 
$[x(s), y(s), z(s)]$, or 2D projections $[x(s), y(s)]$ 
(Malanushenko \etal, 2009, 2012). The method of Malanushenko \etal, (2012) 
employs a Grad-Rubin type NLFFF code (Grad and Rubin, 1958) that fits a LFFF 
with a local $\alpha$-value to each coronal loop and then iteratively relaxes 
to the closest NLFFF solution, 
while the method of Aschwanden \etal, (2012a) uses an approximative analytical 
solution of a force-free and divergence-free field, parameterized by a number
of buried magnetic charges that have a variable twist around their vertical 
axis, and is forward-fitted to observed loop coordinates. The latter
method is numerically quite efficient and achieves a factor of two better
agreement in the misalignment angle ($\mu=14^\circ-19^\circ$) than standard
NLFFF codes using magnetic vector data. However, the major limitation of the
latter method is the availability of solar stereoscopic data, which restricts
the method to the beginning of the STEREO mission (i.e., the year of 2007),
when STEREO had a small spacecraft separation angle that is suitable for 
stereoscopy (Aschwanden \etal, 2012b). 

Hence, the application of NLFFF magnetic modeling using coronal constraints
could be considerably enhanced, if the methodical restriction to coronal
3D data, as it can be provided only by true stereoscopic measurements, could
be relaxed to 2D data, which could be furnished by any high-resolution EUV
imager, such as from the SoHO/EIT, TRACE, and SDO/AIA missions. This
generalization is exactly the purpose of the present study. We develop a
modified code that requires only a line-of-sight magnetogram and 
a high-resolution EUV image, where we trace 2D loop coordinates to
constrain the forward-fitting of the analytical NLFFF code described in
Aschwanden (2012a), and compare the results with those obtained from
stereoscopically triangulated 3D loop coordinates (described in Aschwanden
\etal, 2012a). Furthermore we test also magnetic forward-fitting to 
automatically
traced 2D loop data, and compare the results with manually traced 2D loop data.
The latter effort brings us closer to the ultimate goal of fully automated 
(NLFFF) magnetic field modeling with widely accessible input data. 

The content of the paper is a follows: the theory of the analytical NLFFF 
forward-fitting code is briefly summarized in Section 2, the numerical code 
is described in Section 3, tests of NLFFF forward-fitting to simulated data
are presented in Section 4, and to stereoscopic and single-image data 
in Section 5, while a discussion of the application is given in Section 6, 
with a summary of the conclusions provided in Section 7. 

\section{          	    ANALYTICAL THEORY 				} 

A nonlinear force-free field (NLFFF) is implicitly defined by Maxwell's
force-free and divergence-free conditions,
\begin{equation}
                {{\bf j} \over c} = {1 \over 4\pi} (\nabla \times {\bf B}) =
                \alpha({\bf x}) {\bf B} \ .
\end{equation}
\begin{equation}
                \nabla \cdot {\bf B} = 0 \ , 
\end{equation}
where $\alpha({\bf x})$ is a scalar function that varies in space,
but is constant along a given field line, and the current density ${\bf j}$
is co-aligned and proportional to the magnetic field ${\bf B}$.
A general solution of Equations (1)-(2) is not available, but numerical 
solutions are computed (see review by Wiegelmann and Sakurai 2012) 
using (i) force-free and divergence-free optimization algorithms 
(Wheatland \etal, 2000; 
Wiegelmann 2004), (ii) evolutionary magneto-frictional methods 
(Yang \etal, 1986; Valori \etal, 2007), or Grad-Rubin-style 
(Grad and Rubin, 1958) current-field iteration methods 
(Amari \etal, 1999, 2006; Wheatland 2006; Wheatland and Regnier 2009;
Malanushenko \etal, 2009).
Numerical NLFFF solutions bear two major problems: (i) every method
based on extrapolation of force-free magnetic field lines from 
photospheric boundary conditions suffers from the inconsistency
of the photospheric boundary conditions with the force-free assumption 
(Metcalf \etal 1995), and (ii) the calculation of a single NLFFF 
solution with conventional numerical codes is so computing-intensive 
that forward-fitting to additional constraints (requiring many 
iteration steps) is unfeasible.
Hence an explicit analytical solution of Equations (1)-(2) would be
extremely useful, which could be computed much faster and be
forward-fitted to coronal loops in force-free domains circumventing
the non-force-free (photospheric) boundary condition. 

An approximate analytical solution of Equations (1)-(2) was recently
calculated (Aschwanden 2012a) that can be expressed by a superposition 
of an arbitrary number of $N_m$ magnetic field components 
${\bf B}_m$, $j=1,...,N_m$, 
\begin{equation}
        {\bf B}({\bf x}) = \sum_{m=1}^{N_m} {\bf B}_m({\bf x}) \ ,
\end{equation}
where each magnetic field component ${\bf B}_m$ can be decomposed
into a radial $B_r$ and an azimuthal field component $B_\varphi$, 
\begin{equation}
        B_r(r, \theta) = B_m \left({d^2 \over r^2}\right)
        {1 \over (1 + b^2 r^2 \sin^2{\theta})} \ ,
\end{equation}
\begin{equation}
        B_\varphi(r, \theta) =
        B_m \left({d^2 \over r^2}\right)
        {b r \sin{\theta} \over (1 + b^2 r^2 \sin^2{\theta})} \ ,
\end{equation}
\begin{equation}
        B_\theta(r, \theta) \approx 0
        \ ,
\end{equation}
\begin{equation}
        \alpha(r, \theta) \approx {2 b \cos{\theta} \over
        (1 + b^2 r^2 \sin^2{\theta})}  \ ,
\end{equation}
where ($r, \varphi, \theta$) are the spherical coordinates of a 
magnetic field component system ($B_m, x_m, y_m, z_m, \alpha_m)$
with a unipolar magnetic charge $B_m$ that is buried at position
($x_m, y_m, z_m)$, has a depth $d=1-(x_m^2+y_m^2+z_m^2)^{1/2}$, 
a vertical twist $\alpha_m$, and $r=[(x-x_m)^2+(y-y_m)^2+(z-z_m)^2]^{1/2}$
is the distance of an arbitrary coronal position $(x,y,z)$ to the
subphotospheric location $(x_m, y_m, z_m)$ of the buried magnetic charge.
The force-free parameter $\alpha$ can also be expressed in terms
of the parameter $b$ (Equation 7), which quantifies the number $N_{twist}$ 
of full twist turns over a (loop) length $L$,
\begin{equation}
        b = {2 \pi N_{twist} \over L} .
\end{equation}
This analytical approximation is divergence-free and force-free
to second-order accuracy in the parameter $(b\ r \sin \theta)$, 
which is proportional to the force-free 
parameter $\alpha$ as defined by Equation (7). This approximate NLFFF
solution is very appropriate for cases with small vertical twist,
but may break down for highly non-potential cases with large
twist or magnetic field domains with strong horizontal twist,
such as near-horizontal filaments or a Gold-Hoyle flux rope
(Gold and Hoyle 1960; Aschwanden 2012a, Appendix A).
In the limit of
vanishing vertical twist ($\alpha \mapsto 0$ or $b \mapsto 0$),
the azimuthal component vanishes, $B_\varphi \mapsto 0$, and the
radial component degenerates to the potential-field solution of
a unipolar magnetic  charge, $B_r \mapsto B_m (d/r)^2$, which
is simply a radial field that points away from the buried charge
and decreases with the square of the distance.
A numerical code that fits this analytical approximation
to a given 3D magnetic field is described and tested in
Aschwanden and Malanushenko (2012) using analytical models. 
Applications to real solar data using stereoscopically triangulated 
coronal loops from the STEREO mission, which supposedly outline 
force-free magnetic field lines in the solar corona, are presented in
Aschwanden \etal, (2012a). 

\section{	NUMERICAL FORWARD-FITTING		} 

The numeric code that fits the approximative analytical NLFFF solution
(Equations 3-7) to a line-of-sight magnetogram $B_z(x,y,z_{phot})$ plus
coronal loop coordinates $[x(s), y(s), z(s)]$ for an ensemble of
stereoscopically triangulated loops in a solar active region is
described in detail and tested in Aschwanden and Malanushenko (2012).
We are using a cartesian coordinate system $(x,y,z)$ with the
origin in the center of the Sun and the plane-of-the-sky is
in the $(x,y)$-plane, while the z-axis is the line-of-sight. 
This allows us to take the curvature of the solar surface into 
full account, in contrast to some other NLFFF codes that 
approximate the solar surface with a flat plane. 

The forward-fitting part of the code consists of two major parts,
(i) the decomposition of buried unipolar magnetic charges from
a line-of-sight magnetogram $B_z(x,y,z_{phot})$ (see Appendix A of
Aschwanden \etal, 2012a), and (ii) iterative optimization of the
nonlinear force-free $\alpha({\bf x})$ parameters by minimizing
the misalignment angles between the loop data and the fitted NLFFF model.
The new approach in this work is the generalization of the 
forward-fitting code from 3D loop coordinates (using stereoscopic
measurements before) to 2D loop coordinates $[x(s), y(s)]$ only, 
which can simply be provided from any high-resolution EUV image without 
requiring stereoscopic views. 

The two different methods are juxtaposed in Fig.~1. In the 3D 
forward-fitting method, the 3D misalignment angle
$\mu_3(i,j), i=1,...,n_s, j=1,...,n_L$ is computed for a number 
of $n_s$ loop segment positions and $n_L$ coronal loops, defined by 
the scalar product that calculates the angle $\mu_3$ between the
3D vectors of the loop direction ${\bf B}^{obs}({\bf x})$ and
the magnetic field direction ${\bf B}^{theo}({\bf x})$ at a given
loop position ${\bf x}(s)=[x(s_{ij}), y(s_{ij}), z(s_{ij})]$,
\begin{equation}
        \mu_3({\bf x}) =
        cos^{-1} \left({ {\bf B}^{theo}({\bf x}) \cdot
        {\bf B}^{obs}({\bf x}) \over
        |{\bf B}^{theo}({\bf x})|\ |{\bf B}^{obs}({\bf x})| }\right) \ .
\end{equation}
This is illustrated in Figure (1), where the loop directions (black arrows)
and magnetic field directions (red arrows) are depicted at three loop
segment positions, for two orthogonal projections (Figure 1 left panels). 
The root-mean-square value of all misalignment
angles for each loop segment $(i)$ and loop $(j)$ is then minimized
in the forward-fitting procedure to find the best NLFFF approximation,
\begin{equation}
	\mu_3 = {1 \over n_s n_L} \left( \sum_{i,j}^{n_s, n_L} 
			[\mu_3({\bf x}_{ij})]^2 \right)^{1/2} \ .
\end{equation}
In addition to the 3D misalignment angle $\mu_3$, we can also define
a 2D misalignment angle $\mu_2$ with the same Equations (9) and (10),
except that the magnetic field vectors ${\bf B}^{theo}({\bf x)}$ and
loop vectors ${\bf B}^{theo}({\bf x})$ are a function of two-dimensional
space coordinates ${\bf x}=(x_{ij}, y_{ij})$, as they are seen in the
2D projection into the $(x,y)$-plane (Figure 1, bottom right panel).

If we have only 2D loop coordinates available (in the case without
stereoscopy), we can only forward-fit the field lines parameterized with 
a NLFFF code by minimizing the 2D misalignment angle $\mu_2$, because
the third space variable $(z_{ij})$ is not available. Our strategy
here is to calculate the 2D misalignment angle $\mu_2$ in each position for
an array of altitudes, $r_i=1 + h_{max} (i / n_h), i=1,...,n_h$, that 
covers a limited altitude range $[1 < r < (1+h_{max})]$ or search volume 
in which we expect coronal loops to be detectable 
(Figure 1, top right panel). For EUV images,
loops are generally detectable within one density scale height, 
for which we use a height range of $[1 < r < 1.15]$ solar radii here.
This yields multiple $(n_h)$ misalignment angles for each 2D loop
position $(x_{ij}, y_{ij})$, which are shown in the $(x-z)$-plane
(Fig.~1 top right panel) and $(x-y)$-plane (Fig.~1 bottom right panel). 
Our strategy is then to estimate the unknown third $z_{ij}$-coordinate
in each 2D position $(x_{ij}, y_{ij})$ from that height $r_i$ that
shows the smallest 2D misalignment angle $\mu_2$, and can then proceed
with the forward-fitting procedure like in the case of 3D stereoscopic
data $(x_{ij}, y_{ij}, z_{ij})$, which is described in detail
in Aschwanden and Malanushenko (2012). 

A side-effect of the generalized
2D method is that the parameter space is enlarged by an additional
dimension, \ie the unknown $z_i$ or $r_i$ coordinate of the altitude
of each loop position $(x_{ij}, y_{ij})$, which in principle increases 
the computation time by a linear factor of the number $n_h$ of altitude
levels. However, we optimized the code by vectorization and by
organizing the optimization of the altitude variables $z_{ij}$ 
(for each loop segment $j$ and loop $i$) and the $\alpha_k$-parameter 
variables (for each magnetic charge $m$) in an interleaved mode, 
so that the computation time reduced by 1-2 orders of magnitude compared
with earlier versions of the code (Aschwanden and Malanushenko 2012;
Aschwanden \etal, 2012a), without loss of accuracy. 

When comparing 3D with 2D misalignment angles, we have to be aware
that the unknown third dimension $z_{ij}$ at every loop position
$(x_{ij}, y_{ij})$ is handled differently in the two methods.
In our new 2D method we optimize the third coordinate $z_{ij}$ by 
minimizing the 2D misalignment angle independently at every 
loop segment position $(x_{ij}, y_{ij})$, interleaved with the optimization
of the nonlinear force-free $\alpha_m$ parameter. In the (stereoscopic)
3D method the third coordinate $z_{ij}$ is used as a fixed constraint
like the observables $(x_{ij}, y_{ij})$. However, we can calculate
the median 2D misalignment angle $\mu_2$ with both methods, while
the 3D misalignment angle $\mu_3$ is only defined for the (stereoscopic)
3D-fit method, but not for the (loop-tracing) 2D-fit method.

\section{	TEST RUNS WITH SIMULATED DATA 		} 

In order to test the numerical convergence behavior, the uniqueness of
the solutions, and the accuracy of the method in terms of misalignment  
angles we test our code first with simulated data. We simulate six cases
that correspond to the same six nonpotential cases presented in Aschwanden 
and Malanushenko (2012; cases \# 7-12), consisting of a unipolar case
(N7; Figure 2 top), a dipolar case (N8; Figure 2 middle), a quadrupolar
case (N9; Figure 2 bottom), and three decapolar cases with 10 randomly
buried magnetic charges each (N10, N11, N12; Figure 3). In each case
we run both the 3D-fitting code (mimicking the availability of
stereoscopic loop data), as well as the 2D-fitting code (corresponding
to loop tracings from a single EUV image without stereoscopic information).
Thus, in the 3D-fitting code the target loops are parameterized with
3D data $[x_i(s), y_i(s), z_i(s)]$, while we ignore the third coordinate
in the 2D-fitting code and fit only the 2D coordinates $[x_i(s), y_i(s)]$
of the target loops. The results are shown in Figures 2 and 3, with the
2D fits in the left-hand panels, and the 3D fits in the right-hand panels.

The convergence of the 2D-code can be judged by comparing with the previously
tested 3D-code (Aschwanden and Malanushenko 2012). We list a summary of the
results in Table 1. The 3D-fits achieve a mean 3D misalignment angle of
$\mu_3=3.6^\circ\pm1.9^\circ$ and a 2D misalignment angle of
$\mu_2=2.0^\circ\pm1.1^\circ$. This is reasonable and slightly better than
the results of an earlier version of the 3D-code 
($\mu_3=5.1^\circ\pm4.3^\circ$ in Table 4 of Aschwanden and Malanushenko 2012).
In comparison, our new 2D-fitting code achieves an even better agreement
with $\mu_2=1.2^\circ\pm0.5^\circ$ for the 2D-misalignment angle $\mu_2$,
while the 3D-misalignment angle $\mu_3$ is not defined for the 2D code 
(due to the lack of line-of-sight coordinates $z$). The better
performance of the 2D code is due to the smaller number of constraints
(i.e., 200 constrains for 2D-loop coordinates of 10 loops with 10 segments,
compared with 300 constraints for the 3D-fit method). The smaller number
of free parameters generally improves the accuracy of the solution.
The uniqueness of the solution can be best expressed by the mean misalignment
angle $\mu_2$. Strictly speaking, a NLFFF solution would only be unique if
the number of free parameters (i.e., the number of magnetic charges $N_m$
in our case) matches the number of constraints (i.e., the number of fitting
positions (i.e., the product of the number of loops $N_{\rm loop}$ times 
the number of fitted segment positions $N_{segm}$, i.e., $N_{fit}=N_{loop}
\times N_{seg}$ in our case). Moreover, the LOS magnetogram is approximated
by a number of magnetic charges $N_m$ that neglects weak magnetic sources,
and there are residuals in the decomposition of magnetic charges that
contribute to the noise or uncertainty and non-uniqueness of the solutions.
Therefore, the uniqueness of a NLFFF solution can best be specified by
an uncertainty measure for each field line, which can be quantified either
by the misalignment angle $\mu_2$ or by a maximum transverse displacement 
of a field line, i.e, $\Delta x \approx (L/2) \tan \mu_2$ for a particular 
field line with full length $L$.

We show in Table 1 also the ratios $E_{NP}/E_P$ of the nonpotential 
to the potential energies for the 3D and 2D fit methods, which agree within an 
accuracy of order $\approx 0.1-2.5 \%$. Note, that the simulated data have
1, 2, 4, and 10 magnetic charges, which corresponds to the number of
free parameters in the fit, while we used the double number of magnetic
charges in the decomposition of the simulated LOS magnetogram, in order
to make the parameterization of the fitted model somewhat different from
the target model. Nevertheless, although the 2D solutions have a high 
accuracy (with a mean misalignment of $\mu_2=1.2^\circ \pm 0.5^\circ$), 
we have to be aware that the simulated data and the forward-fitting code 
use the same parameterization of nonlinear $\alpha$-parameters, which 
warrants a higher accurcay in forward-fitting than real data with a 
unknown parameterization. Hence, we test the code with real solar data 
in the following section. 

\section{	OBSERVATIONS AND RESULTS 		} 

\subsection{Observations}

For testing the feasibility, fidelity, and accuracy of the new
analytical NLFFF forward-fitting code based on 2D tracing of loops
(rather than 3D stereoscopy) we are using the same observations for
which either stereoscopic 3D reconstruction has been attempted earlier
(Aschwanden \etal, 2008b,c, 2009, 2012a; Sandman \etal, 2009; DeRosa \etal, 
2009; Aschwanden and Sandman 2010; Sandman and Aschwanden 2011; 
Aschwanden 2012a),
such as for active regions observed with STEREO on 2007 April 30, May 9,
May 19, and December 11, or where 2D loop tracing was performed and documented,
such as for an active region observed on 1998 May 19 with TRACE (Aschwanden 
\etal, 2008a). The active region numbers, observing times, spacecraft
separation angles, number of traced loops, and maximum magnetic field
strengths of these observations are listed in Table 2. 
In all cases we used line-of-sight magnetograms $B_z(x,y,z_{phot})$ from
the Michelson Doppler Imager (MDI; Scherrer \etal, 1995) on board the
{\sl Solar and Heliospheric Observatory (SOHO)}, while EUV images
were used either from the {\sl Transition Region And Coronal Explorer} (TRACE;
Handy \etal, 1999), or from the {\sl Extreme Ultraviolet Imager} (EUVI; 
W\"ulser \etal, 2004) onboard the STEREO spacecraft A(head) and B(ehind).

We show the results of the NLFFF forward-fitting of the six active regions
in Figures 4 to 9, all in the same format, which includes the decomposed
line-of-sight magnetogram of SoHO/MDI (grey scale in center of Figures 4 to 9),
the stereoscopically triangulated or visually traced loops (blue curves
in Figures 4 to 9), and the best-fit magnetic field lines (red curves for
the segments covered by the observed loops, and in orange color for 
complementary
loops parts (although truncated at a height of 0.15 solar radii). The
orthogonal projections of the best-fit magnetic field lines are also
shown in the right-hand and top panels of Figures 4 to 9, as well as the
histograms of 2D $(\mu_2)$ and 3D ($\mu_3$) misalignment angles (in
bottom panels of Figures 4 to 9). The median misalignment angles
$\mu_2$ and $\mu_3$ are also listed in Table 3, for both the previous
stereoscopic reconstructions (Aschwanden \etal, 2012a; Aschwanden 2012b),
marked with "STEREO" in Table 3, and based on 2D loop tracing in the 
present study, marked with "Tracing" in Table 3. 

The active region A (2007 April 30; Figure 4) shows a lack of stereoscopically
triangulated loops in the core of the active region (due to the high level
of confusion for loop tracing over the ``mossy'' regions), where the highest
shear and degree of non-potentiality is expected, and thus deprives us from
measuring the largest amount of free magnetic energy, while standard NLFFF
codes have stronger constraints in these core regions. In all 5 active 
regions we see sunspots with strong magnetic fields, but since we limit
our NLFFF solutions to an altitude range of $\lapprox 0.15$ solar radii,
we cannot see whether the diverging field lines above the sunspots are
open or closed field lines. In principle we could display our NLFFF
solutions to larger altitudes to diagnose where open and closed field
regions are, but the accuracy of reconstructed field lines is expected
to decrease with height with our method, especially because of the
second-order approximation that can represent helical twist well for
vertical segments of loops (near the surface), but not so well for
horizontal segments of loops (in large altitudes). 
 
\subsection{Misalignment Angle Statistics}

We see in Table 3 that the 3D misalignment angles $\mu_3$ have 
a mean value of $\mu_3=19^\circ\pm3^\circ$ for the stereoscopy method. 
If we compare the 2D misalignment angles $\mu_2$ between the two methods
in Table 3, we see that the STEREO method has a somewhat larger mean, 
$\mu=4.0^\circ\pm1.8^\circ$, than the 2D loop tracing method, with
$\mu=2.7^\circ\pm1.3^\circ$. This is an effect of the optimization of the 
2D misalignment angles in the 2D loop tracing method, where the third
coordinate is a free variable and thus has a larger flexibility to find
an appropriate model field line with a small 2D misalignment angle,
in contrast to the stereoscopic method, where the third coordinate of
every observed loop is entirely fixed and leaves less room in the
minimization of the misalignment angles between observed loops and 
theoretical field models. Ideally, if stereoscopy would work perfectly,
the third coordinate should be sufficiently accurate so that the best
field solution can be found easier with fewer free parameters. However,
reality apparently reveals that there is a significant stereoscopic
error that hinders optimum field fitting, which is non-existent in the
2D forward-fitting situation. Actually, from the two stereoscopic
misalignment angles we can estimate the stereoscopic error $\sigma_{SE}$, 
assuming isotropic errors. Thus, defining the 2D misalignment angle 
as $\mu_2^2=\sigma_x^2+\sigma_y^2$, and the 3D misalignment angle as 
$\mu_3^2=\sigma_x^2+\sigma_y^2+\sigma_z^2+\sigma_{SE}^2$, with
isotropic errors $\sigma_x=\sigma_y=\sigma_z$, we expect
\begin{equation}
	\sigma_{SE}^2 = \mu_3^2 - \left({3 \over 2}\right) \mu_2^2 \ ,
\end{equation} 
which yields a stereoscopic error of $\sigma_{SE} \approx 20^\circ$. 
This is somewhat larger than estimated earlier from the parallelity
of stereoscopically triangulated loops with close spatial proximity,
which amounted to $\sigma_{SE} \approx 7^\circ=12^\circ$ (Aschwanden
and Sandman 2010). Thus both estimates assess a substantial value to
the stereoscopic error that exceeds the accuracy of the best-fit NLFFF  
solution based on 2D loop tracing ($\mu_2 = 2.7^\circ\pm1.3^\circ$) 
by far. Nevertheless, both best-fit misalignment angles $\mu_2$ 
are consistent with each other for the two forward-fitting
methods. This forward-fitting experiment thus demonstrates that 
{\sl we obtain equally accurate NLFFF fits to coronal data with or without
stereoscopy}, and thus makes the new method extremely useful. 

\subsection{Spatial Resolution of Instruments}

For the 2D loop tracing method of stereoscopic data (cases A, B, C, and D
in Table 1) we just ignored information on the $z$-coordinate
of loops in the 2D forward-fitting algorithm. For cases E and F,
we compare visual tracing of loops (case E) with automated loop tracing 
(case F). Interestingly, we find the most accurate forward-fit for case E,
which has a 2D misalignment error of only $\mu_2=1.2^\circ$ (Figure 8), 
which may have resulted from the higher spatial resolution of loop tracing
using TRACE data (with a pixel size of $0.5\arcsec$) in case E, 
compared with the three times lower resolution of STEREO (with a 
pixel size of $1.6\arcsec$) in the cases A, B, C, and D. 

\subsection{Visual versus Automated Loop Detection}

Comparing visually traced (Figure 8) versus automatically traced loops 
(Figure 9), we note that the automated tracing leads to a less
accurate NLFFF fit, with $\mu_3=3.3^\circ$ versus $\mu_3=1.2^\circ$
for visual tracing. Apparently, the automated loop tracing method
(Aschwanden \etal, 2008a) can easily be mis-guided or side-tracked
by about $\Delta \mu \approx 2^\circ$ onto near-cospatial loops 
with similar coherent large curvature. Comparing individual field
lines in Figure 8 with Figure 9, we see also that the automated loop
tracing algorithm produces a number of short loop segments with large
misalignment angles, which are obviously a weakness of the automated
tracing code, resulting into a larger misalignment error for the
fitted NLFFF solution (which is a factor of $\approx 3$ larger
in the average for this case).

\subsection{Magnetic Field Strengths}

In Figure 10 we show scatterplots of the magnetic field strengths $B_3$ 
retrieved at the photospheric level for the stereoscopic 3D method versus
the field strength $B_2$ obtained from the 2D loop tracing method. 
The analytical NLFFF forward-fitting algorithm
starts first a decomposition of point-like buried magnetic charges from
the line-of-sight component $B_z(x,y,z_{phot})$, but has no constraints
for the transverse components $B_x$ and $B_y$, except for the coronal
loop coordinates. Thus, the directions of coronal loops near the 
footpoints will determine the transverse components $B_x$ and $B_y$
in the photosphere. The scatterplots of $B_2$ versus $B_3$ in Figure 10 
show that the smallest scatter between the two methods appears for 
active region B, which is the one closest to a potential field (although
we are not able to trace and triangulate loops in the core of the active 
region, where supposedly the highest level of non-potentiality occurs).
The linear regression fits show a good
correspondence in the order of $\approx 0.1\%-1.0\%$ between the two methods, 
which results from different fitting criteria of the nonpotential field
components. A ratio of $B_\varphi/B_r \approx 10\%$ in the azimuthal field 
(or nonpotential) field component $B_\varphi$ (Equation 5) with respect 
to the potential field component $B_r$ (Equation 4) would result into a 
change $B \approx \sqrt{(1+0.1^2)}$ of $\approx 0.5\%$ in the magnetic field 
strength.

\subsection{Magnetic Energies}

The free magnetic energy, which is the difference between the nonpotential
and the potential field energy, integrated over the spatial volume of an
active region, has been calculated for the stereoscopic forward-fitting
method in Aschwanden 2012b (Table 1 therein). Here we calculate these
quantities also for the 2D loop tracing method, both methods being juxtaposed
in Table 3 and in Figure 11. The absolute values of the potential field
energy $E_P$ agree within a factor of $1.04\pm0.02$ between the two methods,
while the free energies show differences of 1\%-2\% of the potential energy.
If we consider the difference in the free energy between two methods as
a measure of a systematic error, we assess an uncertainty of 
$E_{free}/E_P-1=0.016\pm0.013$, or about $\pm 1.5\%$.

\section{	DISCUSSION 				} 

\subsection{Assessment of Various NLFFF Methods}

Quantitative comparisons of various nonlinear force-free field (NLFFF)
calculation methods applied to coronal volumes that encompass an active
region included optimizational, magneto-frictional, Grad-Rubin based,
and Green's function based models (Schrijver \etal, 2006; 2008; DeRosa
\etal, 2009). A critical assessment of 11 
NLFFF methods revealed significant differences in the nonpotential 
magnetic energy and in the degree of misalignment ($\mu_3 \approx
24^\circ-44^\circ$) with respect to stereoscopically triangulated
coronal loops. The chief problem responsible for these discrepancies
were identified in terms of the non-force-freeness of the lower 
(photospheric) boundary, too small boundary areas, and uncertainties of the
boundary data (DeRosa \etal, 2009). Obviously, the non-force-freeness of the
photosphere can only be circumvented by incorporating coronal magnetic field
geometries into NLFFF extrapolations, such as the information from
stereoscopically triangulated loops.  

Implementation of coronal magnetic field data into NLFFF codes is not
straightforward, because most of the conventional NLFFF codes are designed
to extrapolate from a lower boundary in upward direction, and thus the
volume-filling coronal information cannot be treated as a boundary problem.
A natural method to match arbitrary constraints that are not necessarily
a boundary problem is a forward-fitting approach, which however, requires
a parameterization of a magnetic field model. Since NLFFF models are
implicitly defined by the two differential equations of force-freeness
and divergence-freeness, an explicit parameterization of a magnetic field is
not trivial. There are essentially three approaches that have been attempted 
so far: (i) Preprocessing of magnetic boundary data by minimizing the 
Lorentz force at multiple spatial scales (Wiegelmann 2004; Jing \etal, 2009);
(ii) linear force-free fitting with subsequent relaxation to a nonlinear 
force-free field (Malanushenko \etal, 2009, 2012); and
(iii) magnetic field parameterization with buried magnetic charges and
forward-fitting of an approximative analytical NLFFF solution
(Aschwanden 2012a,b; Aschwanden and Malanushenko 2012).   
The third method requires stereoscopically triangulated 3D coordinates
of coronal loops, which have been successfully modeled in four different
active regions, improving the misalignment angles between the observed
loops and a potential field ($\mu_3=19^\circ-46^\circ$) by a factor of
about two, when fitted with the analytical NLFFF solution ($\mu_3=
14^\circ-19^\circ$; Aschwanden \etal, 2012a). However, although this
method provides a more realistic NLFFF model, the main restriction is the
availability of STEREO data, as well as the inferior spatial resolution of
STEREO/EUVI compared with other EUV imagers (e.g., TRACE, or SDO/AIA).
In order to circumvent this restriction we developed a generalized code
in this study that requires only a high-resolution EUV image, from
which a sufficient large number of coronal loops can be traced in two
dimensions, plus a line-of-sight magnetogram, which were essentially
always available since the lauch of the SOHO mission in 1995.  
A fortunate outcome of this study is that the reduction of 3D information
from coronal loops to 2D coordinates does not handicap the accuracy of
a NLFFF forward-fitting method, as we demonstrated here. One potential
limitation of forward-fitting of analytical NLFFF approximations may be
the accuracy for strong nonpotential cases (say near flaring times), since
our analytical NLFFF approximation is only accurate to second order of the
force-free $\alpha$-parameter. Another caveat may be the universality of
the analytical NLFFF parameterization. Our approximation is designed to
model azimuthal (rotational) twist around a vertical axis with respect to the solar
surface, which corresponds to electric currents flowing in vertical
direction. This geometry may not be appropriate for horizontally twisted
structures, such as horizontally extended filaments. For such cases, a more
general NLFFF solution could be desirable. In this respect, more general
NLFFF solutions such as currently developed by Malanushenko \etal, (2009;
2012) may provide new tools, after they will be generalized for a spherical
solar surface and optimized for computational speed. Nevertheless, our
analytical NLFFF forward-fitting algorithm can always be used to find a
quick first approximation of a NLFFF solution, which then could be refined
by alternative (more time-consuming) NLFFF codes. 

\subsection{Computational speed}

Note that our forward-fitting code calculates one trial magnetic field 
configuration in about 
0.01 s, while some $10^4-10^5$ iterations are needed for the convergence
of an approximate NLFFF solution, accumulating to a few minutes total 
computation time.  Our forward-fitting code computes a solution
to an active region with an average computation time of $\approx 10$ minutes
(see CPU times in Table 3) on a Mac OS X, 2 $\times$ 3.2 GHz Quad-Core 
Intel Xeon, 32 GB memory, 800 MHz DDR2 FB-DIMM computer. Simple cases
that can be represented with $N)m \lapprox 20$ magnetic charges, require
only computation times of order $\lapprox 10$ s (see CPU times in Table 1). 

\section{		CONCLUSIONS 				} 

We developed an analytical NLFFF forward-fitting code that requires only
the input of a line-of-sight magnetogram $B_z(x,y,z_{phot})$ and a set
of 2D loop tracing coordinates $(x_i, y_i)$ that can be obtained from any
high-resolution EUV image. This is the first attempt to measure the
coronal magnetic field based on directly observed 2D images alone, 
while we needed stereoscopically triangulated 3D loop coordinates 
$(x_i, y_i, z_i)$ from STEREO in previous studies. There exists only 
one other study (to our knowledge) that heads into the same direction 
of forward-fitting a NLFFF model to 2D data (Malanushenko \etal, 2012), 
using simulated 2D data based on analytical NLFFF solutions (from Low 
and Lou 1990) or previous NLFFF modeling (from Schrijver \etal, 2008). 

We forward-fitted analytical NLFFF approximations to four active regions
(cases A, B, C, D) using 2D loop tracings only and compared them with
previous NLFFF fits using (stereoscopic) 3D loop coordinates. Furthermore
we forward-fitted analytical NLFFF approximations to another active region
using visual 2D loop tracings (case E) or automatically traced 2D loop
coordinates (case F). Our findings of these exercises are the following:

\begin{enumerate}
\item{Our forward-fitting experiment with two different methods
	demonstrated that a nonlinear force-free magnetic field (NLFFF) 
	solution ${\bf B}({\bf x})$ can be obtained with equal accuracy
	{\sl with or without stereoscopy}. This result relinquishes the
	necessity of STEREO data for future magnetic modeling of
	active regions on the solar disk, but the availability of 
	suitable STEREO data was crucial to establish this result.}
\item{The accuracy of a forward-fitted NLFFF approximation that includes
	vertical currents (with twisted azimuthal magnetic field components)
	matches coronal loops observed in EUV with a median 2D
	misalignment angle of $\mu_2=2.7^\circ\pm1.3^\circ$, while
	stereoscopic 3D data exhibit a commensurable 2D misalignment 
	angle, but a substantially larger 3D misalignment angle
	($\mu_3=19^\circ\pm3^\circ$), which implies stereoscopic
	measurement errors in the order of $\sigma_{SE}\approx 20^\circ$.
	These substantial stereoscopic measurement errors lead to less
	accurate NLFFF fits than 2D loop tracings with unconstrained
	line-of-sight positions.}
\item{2D loop tracings in high resolution images (\ie $0.5\arcsec$ pixels
	with TRACE) lead to more accurate NLFFF fits (with a misalignment
	angle of $\mu_2=1.2^\circ$ in case E) than images with lower	
	spatial resolution (\ie $1.6\arcsec$ pixels with STEREO),
	yielding $\mu_2=4.0^\circ\pm1.8^\circ$ for the cases A, B, C, and D.}
\item{Visually (or manually) traced 2D loop coordinates appear to be 
	still superior to the best automated loop tracing algorithms,
	yielding a misalignment angle of $\mu_2=1.2^\circ$ (in case E)
	versus $\mu_2=3.3^\circ$ (in case F).}
\item{Magnetic field strengths $B$ of best-fit NLFFF approximations
	are retrieved with an accuracy of $\approx 0.1\%-1.0\%$, 
	comparing the 2D loop tracing method with the stereoscopic
	3D triangulation method.}
\item{Magnetic energies differ by a factor of $\approx 4\%\pm2\%$ between
	the two methods, while the free energy has a systematic error of
	$\approx \pm 1.5\%$ (of the total magnetic energy) between the two
	methods, which is about an order of magnitude smaller than found
	between other (standard) NLFFF codes.}
\item{The computational speed of our NLFFF code allows the computation
	of a space-filling magnetic field configuration of an active region
	in about 0.01 s, while forward-fitting to a set of coronal loops
	is feasible with about $10^4-10^5$ iterations, requiring a total
	computation time of a few minutes.}
\end{enumerate}

These results demonstrate clearly that we can perform accurate NLFFF 
magnetic modeling based on 2D loop tracings, which will relinquish the
need of STEREO data in future, at least for active regions near the
solar disk center. The analyzed active regions extended up to 0.65
solar radii away from disk center, so in principle we can perform
NLFFF modeling for at least half of the number of active regions
observed on the solar disk, especially since the parameterization
of our analytical NLFFF approximation takes the sphericity of the
solar surface fully into account (which is not the case in most other
NLFFF codes). For future developments we expect that other NLFFF codes
could implement minimization of the misalignment angle $\mu_2$ with
coronal loops in parallel to the optimization of force-freeness 
and divergence-freeness, rather than by preprocessing of vector boundary data.
Further improvements in automated 2D loop tracings could replace 
visual/manual tracing methods and this way render NLFFF modeling 
with coronal constraints in a fully automated way. 
	
\acknowledgements
The author appreciates the constructive comments by an anonymous referee 
and helpful discussions with Allen Gary, Anna Malanushenko,
Marc DeRosa, and Karel Schrijver. Part of the work was supported by
NASA contract NNG 04EA00C of the SDO/AIA instrument and
the NASA STEREO mission under NRL contract N00173-02-C-2035.


\clearpage


\begin{deluxetable}{lrrrrrr}
\tablecaption{Median 2D and 3D misalignment angles $\mu_2$ and $\mu_3$ 
and free energy ratios $E_{NP}/E_P$ for six simulated magnetic field
configurations (see Figures 2 and 3), using the 3D-fit code suitable
for stereoscopic data (marked with "STEREO"), and using the 2D-fit code
suitable for 2D loop data (marked with "Tracing"). The computation time 
(CPU) is listed in seconds.}
\tablewidth{0pt}
\tablehead{
\colhead{Data set}&
\colhead{CPU (s)}&
\colhead{STEREO}&
\colhead{STEREO}&
\colhead{Tracing}&
\colhead{STEREO}&
\colhead{Tracing}\\
\colhead{}&
\colhead{}&
\colhead{$\mu_3$}&
\colhead{$\mu_2$}&
\colhead{$\mu_2$}&
\colhead{$E_{NP}/E_P$}&
\colhead{$E_{NP}/E_P$}}
\startdata
N7	&  0.3 & $0.6^\circ$ & $0.1^\circ$ & $0.7^\circ$ & 1.009 & 1.034 \\
N8	&  2.1 & $2.2^\circ$ & $1.4^\circ$ & $0.6^\circ$ & 1.010 & 1.009 \\	
N9	&  3.0 & $3.7^\circ$ & $2.4^\circ$ & $1.4^\circ$ & 1.016 & 1.015 \\
N10	& 16.5 & $5.0^\circ$ & $3.2^\circ$ & $1.6^\circ$ & 1.055 & 1.054 \\
N11	& 12.5 & $5.5^\circ$ & $2.2^\circ$ & $1.8^\circ$ & 1.066 & 1.050 \\
N12	& 17.8 & $4.6^\circ$ & $2.4^\circ$ & $1.1^\circ$ & 1.140 & 1.094 \\
        &      &             &             &             &               \\      
Mean  	& $8.7\pm7.8$ & $3.6\pm1.9$ & $2.0\pm1.1$ & $1.2\pm0.5$ &  &       \\  
\enddata
\end{deluxetable}

\begin{deluxetable}{lllllrcc}
\tabletypesize{\footnotesize}
\tablecaption{Data selection of four Active Regions observed with SOHO/MDI
and STEREO/EUVI or TRACE.}
\tablewidth{0pt}
\tablehead{
\colhead{Case}&
\colhead{Active}&
\colhead{Observing}&
\colhead{Observing}&
\colhead{Observing}&
\colhead{Spacecraft}&
\colhead{Number}&
\colhead{Magnetic}\\
\colhead{}&
\colhead{Region}&
\colhead{date}&
\colhead{time EUV}&
\colhead{time MDI}&
\colhead{separation}&
\colhead{of EUVI}&
\colhead{field strength}\\
\colhead{}&
\colhead{}&
\colhead{}&
\colhead{(UT)}&
\colhead{(UT)}&
\colhead{angle (deg)}&
\colhead{loops}&
\colhead{B(G)}}
\startdata
A& 10953 (S05E20) &2007-Apr-30 &23:00-23:20 & 22:24 &STEREO 6.0$^\circ$  &200    &[-3134,+1425]\\
B& 10955 (S09E24) &2007-May-9  &20:30-20:50 & 20:47 &STEREO 7.1$^\circ$  &70     &[-2396,+1926]\\
C& 10953 (N03W03) &2007-May-19 &12:40-13:00 & 12:47 &STEREO 8.6$^\circ$  &100    &[-2056,+2307]\\
D& 10978 (S09E06) &2007-Dec-11 &16:30-16:50 & 14:23 &STEREO 42.7$^\circ$ &87     &[-2270,+2037]\\
E&  8222 (N22W30) &1998-May-19 &22:21-22:22 & 20:48 &TRACE (manu)        &201    &[-1787,+1200]\\
F&  8222 (N22W30) &1998-May-19 &22:21-22:22 & 20:48 &TRACE (auto)        &222    &[-1787,+1200]\\
\enddata
\end{deluxetable}

\begin{deluxetable}{lrrrrrr}
\tablecaption{Median 2D and 3D misalignment angles $\mu_2$ and $\mu_3$ 
and free energy ratios $E_{NP}/E_P$ for six 
active regions, including four stereoscopically triangulated loops observed 
with STEREO (A-D) and two cases with traced loops observed with TRACE (E-F). 
All cases have also been modeled by forward-fitting of the NLFFF model to
the traced 2D loop coordinates (columns marked with ``Tracing''). 
The computation time (CPU) is listed in seconds.}
\tablewidth{0pt}
\tablehead{
\colhead{Data set}&
\colhead{CPU (s)}&
\colhead{STEREO}&
\colhead{STEREO}&
\colhead{Tracing}&
\colhead{STEREO}&
\colhead{Tracing}\\
\colhead{}&
\colhead{}&
\colhead{$\mu_3$}&
\colhead{$\mu_2$}&
\colhead{$\mu_2$}&
\colhead{$E_{NP}/E_P$}&
\colhead{$E_{NP}/E_P$}}
\startdata
A) 2007-Apr-30	      &     1103 & $21.4^\circ$  & $4.6^\circ$ & $4.4^\circ$ & 1.006 & 1.007 \\
B) 2007-May-9	      &      208 & $17.9^\circ$  & $3.7^\circ$ & $2.4^\circ$ & 1.023 & 1.009 \\ 
C) 2007-May-19        &      415 & $22.1^\circ$  & $5.9^\circ$ & $3.6^\circ$ & 1.085 & 1.053 \\
D) 2007-Dec-11	      &      390 & $14.7^\circ$  & $1.6^\circ$ & $1.5^\circ$ & 1.044 & 1.026 \\ 
E) 1998-May-19 (manu) &      631 &               &             & $1.2^\circ$ \\
F) 1998-May-19 (auto) &      915 &               &             & $3.3^\circ$ \\ 
		      &          &               &             &             &             \\
(A-F) Mean            &$610\pm342$&$19^\circ\pm3^\circ$    
                      		 & $4.0^\circ\pm1.8^\circ$ & $2.7^\circ\pm1.3^\circ$        \\
\enddata
\end{deluxetable}

\clearpage


\begin{figure}
\plotone{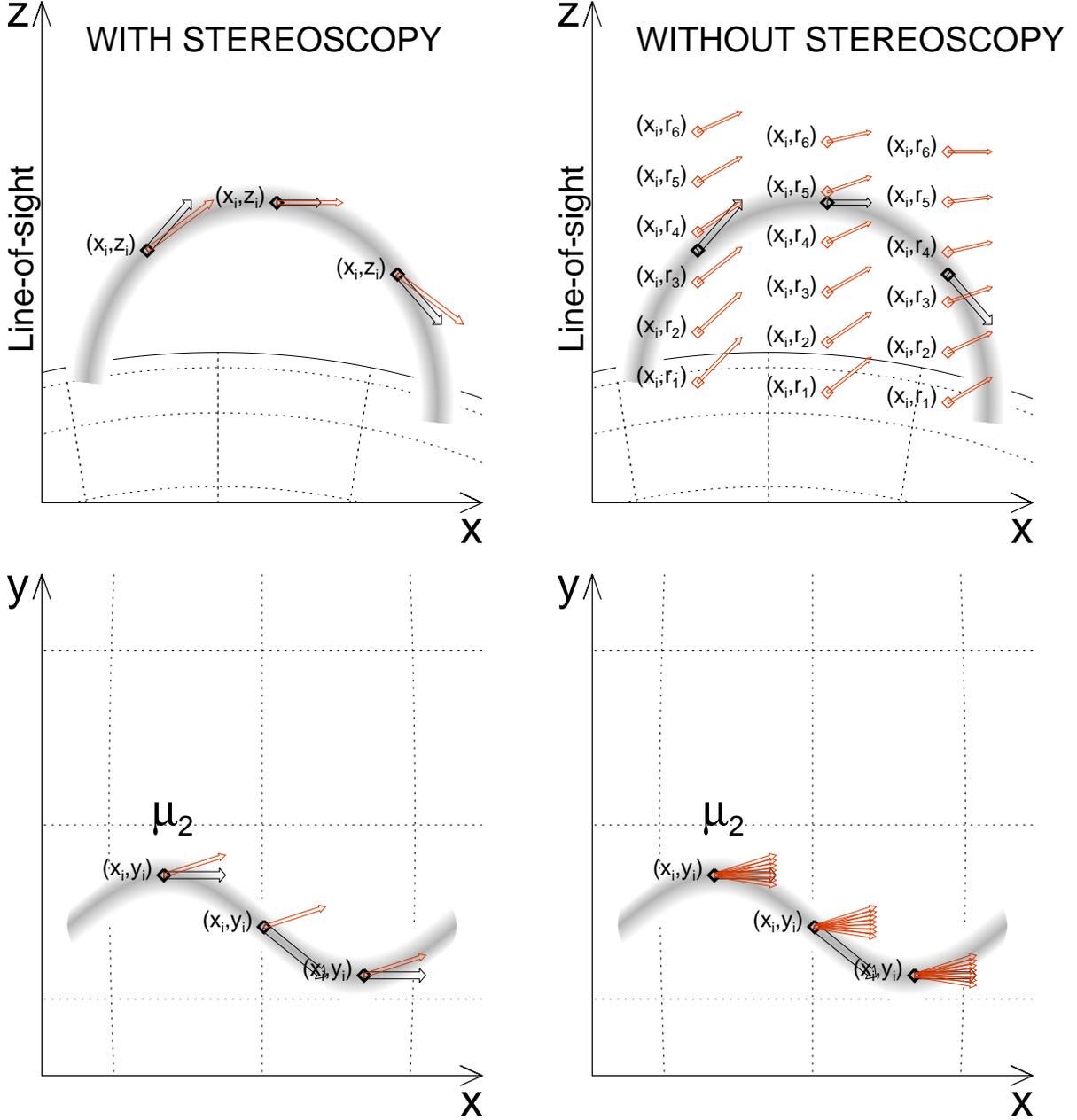}
\caption{A diagram of the optimization parameters in the NLFFF 
forward-fitting method of the magnetic field to observed coronal loops 
is shown for the
case of stereoscopic triangulation (left side), as well as for the case
of 2D loop tracing without stereoscopy (right side). An observed loop
(greyscale) is shown in the plane-of-sky $(x,y)$-plane (lower panels)
and along the line-of-sight in the $(x,z)$-plane, as triangulated with
stereoscopy. The black arrows demarcate the loop directions, while the red
arrows indicate the field direction of the fitted magnetic
field $B(x_i, y_i, z_i)$ at loop positions $(x_i, y_i, z_i)$. In the
case without stereoscopy (right side), an array of field directions
at altitudes $r_1, ..., r_6$ are calculated at projected loop positions
$(x_i, y_i)$. The forward-fitting minimizes the misalignment angle $\mu_2$
between the field directions (red arrows) and the 2D loop
directions (black arrows).}
\end{figure}

\begin{figure}
\plotone{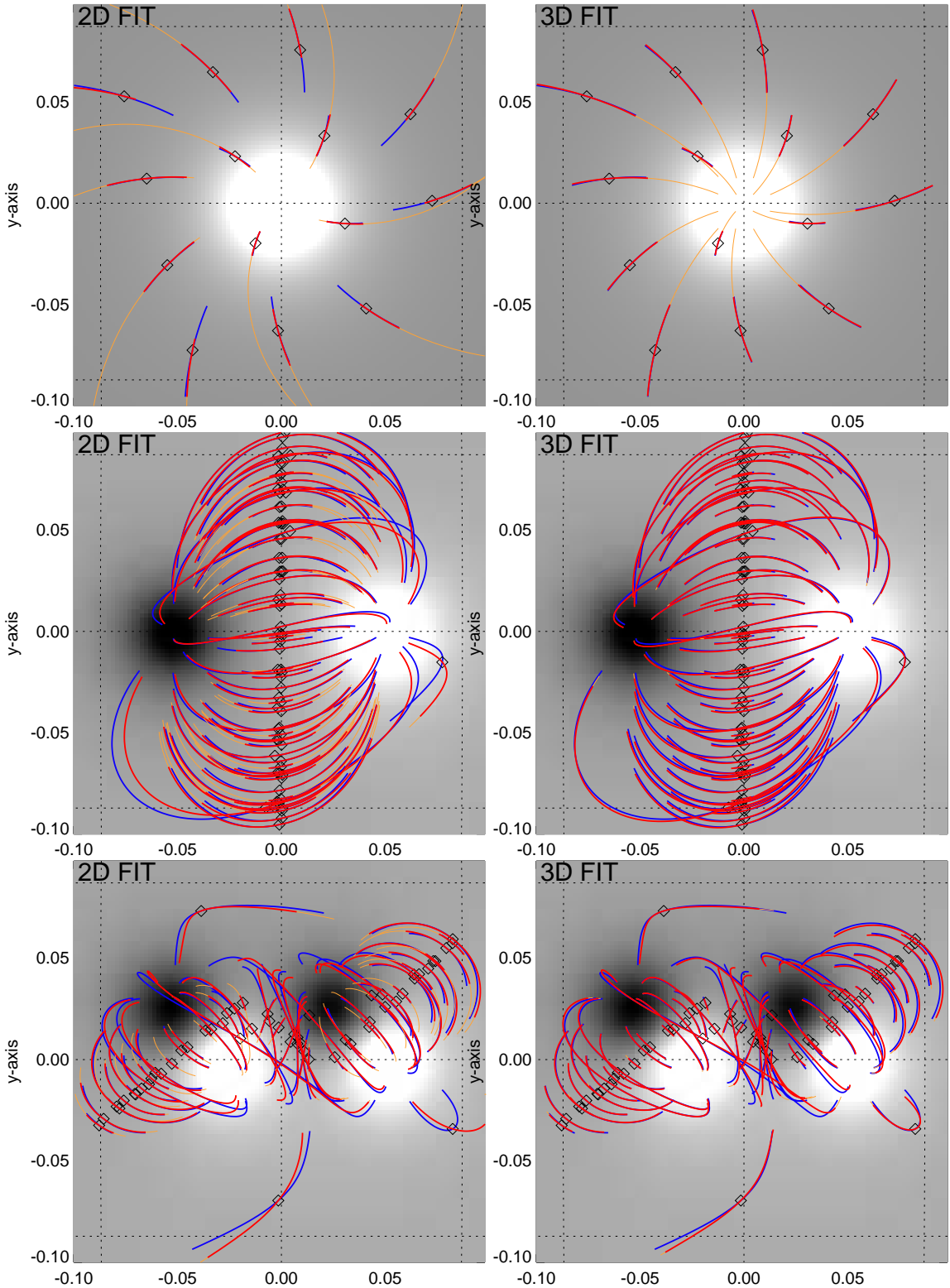}
\caption{Forward-fitting of the NLFFF model to simulated data of
a unipolar (top; case N7), a dipolar (middle; case N8), and quadrupolar 
magnetic configuration (bottom; Case N9), using the 2D-coordinates of 
simulated loops only (left panels)
or the full 3D-coordinates of the simulated loops (right panels).
The blue curves indicate the target loops, the red curves the best-fit 
field lines, and the grey-scale the simulated line-of-sight magnetogram.
The cases N7, N8, and N9 are identical to the simulated cases shown
in Figure (2) of Aschwanden (2012).}
\end{figure}

\begin{figure}
\plotone{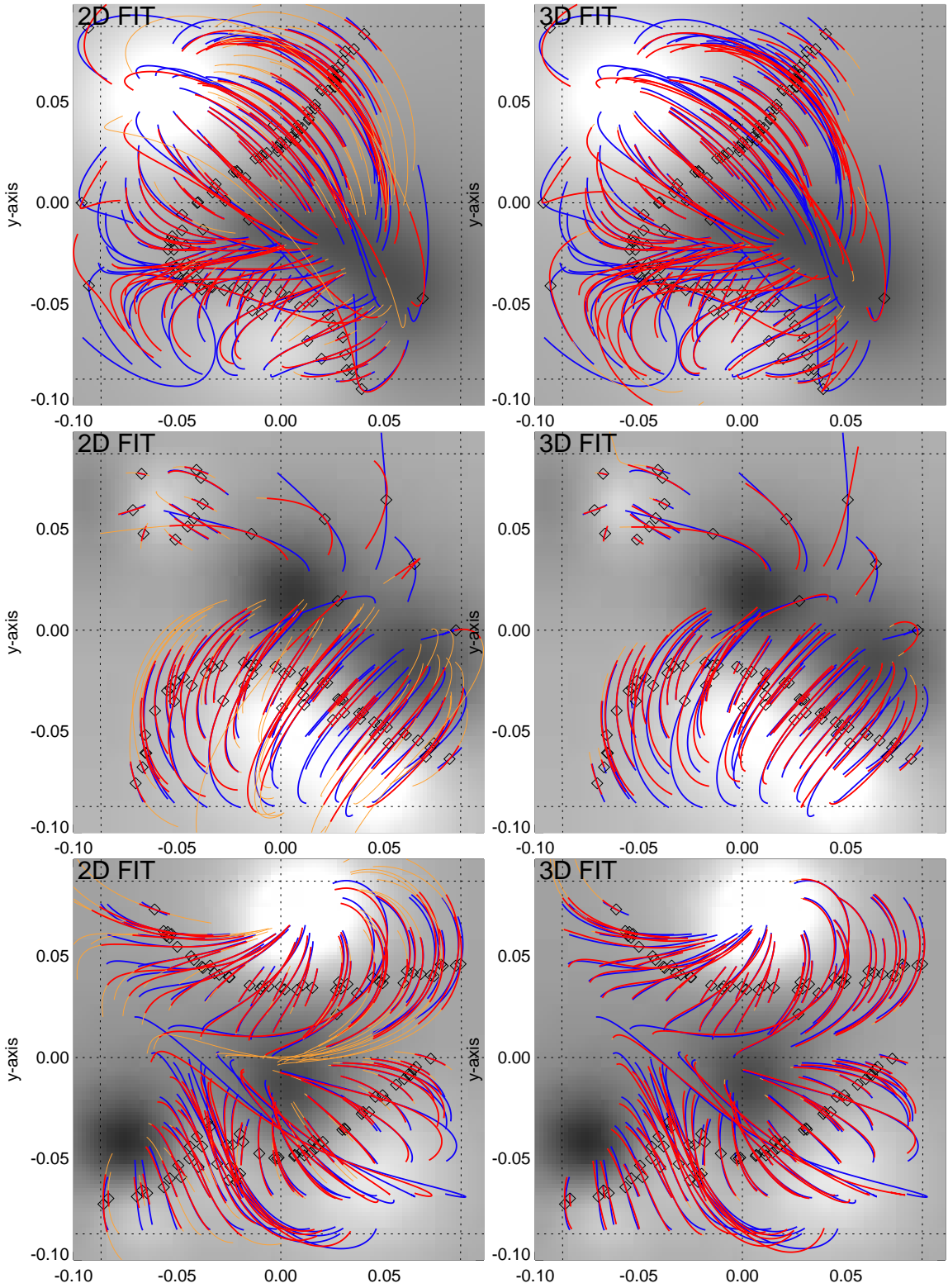}
\caption{Forward-fitting of the NLFFF model to three simulated data cases
with decapolar magnetic configurations (Cases N10, N11, N12, which are identical
to the cases shown in Figure (3) of Aschwanden, 2012), using the 2D-coordinates 
of simulated loops only (left panels) or the full 3D-coordinates of the simulated 
loops (right panels). The blue curves indicate the target loops, the red curves 
the best-fit field lines, and the grey-scale the simulated line-of-sight 
magnetograms.}
\end{figure}

\begin{figure}
\plotone{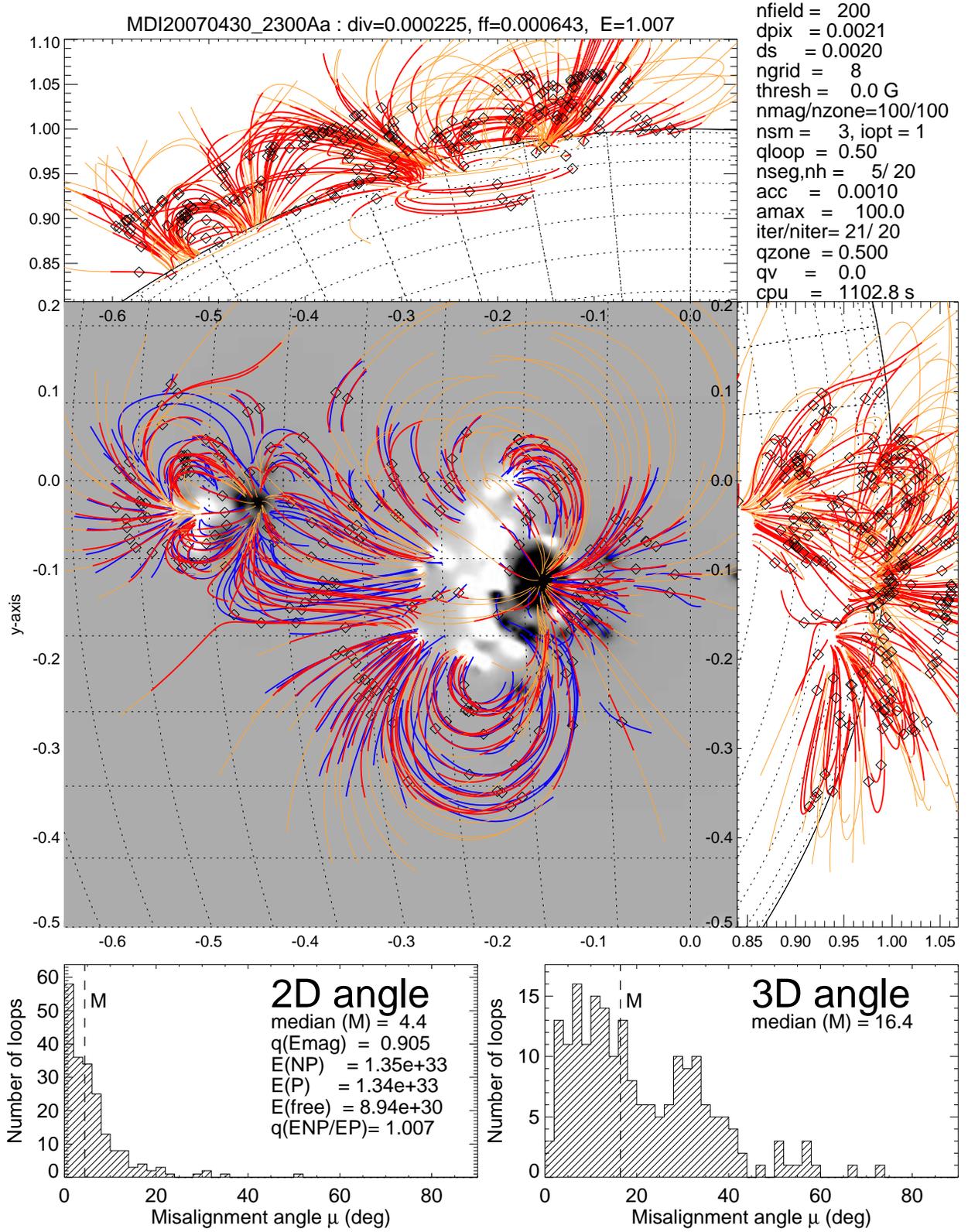}
\caption{Forward-fitting of the NLFFF model to active region 
A (2007 April 30), using STEREO data: 
SoHO/MDI line-of-sight magnetogram (greyscale image),
triangulated STEREO loops (blue curves), and best-fit magnetic field
segments (red) with field line extensions (orange), shown in three
orthogonal projections, and distribution of misalignment angles (bottom).}
\end{figure}

\begin{figure}
\plotone{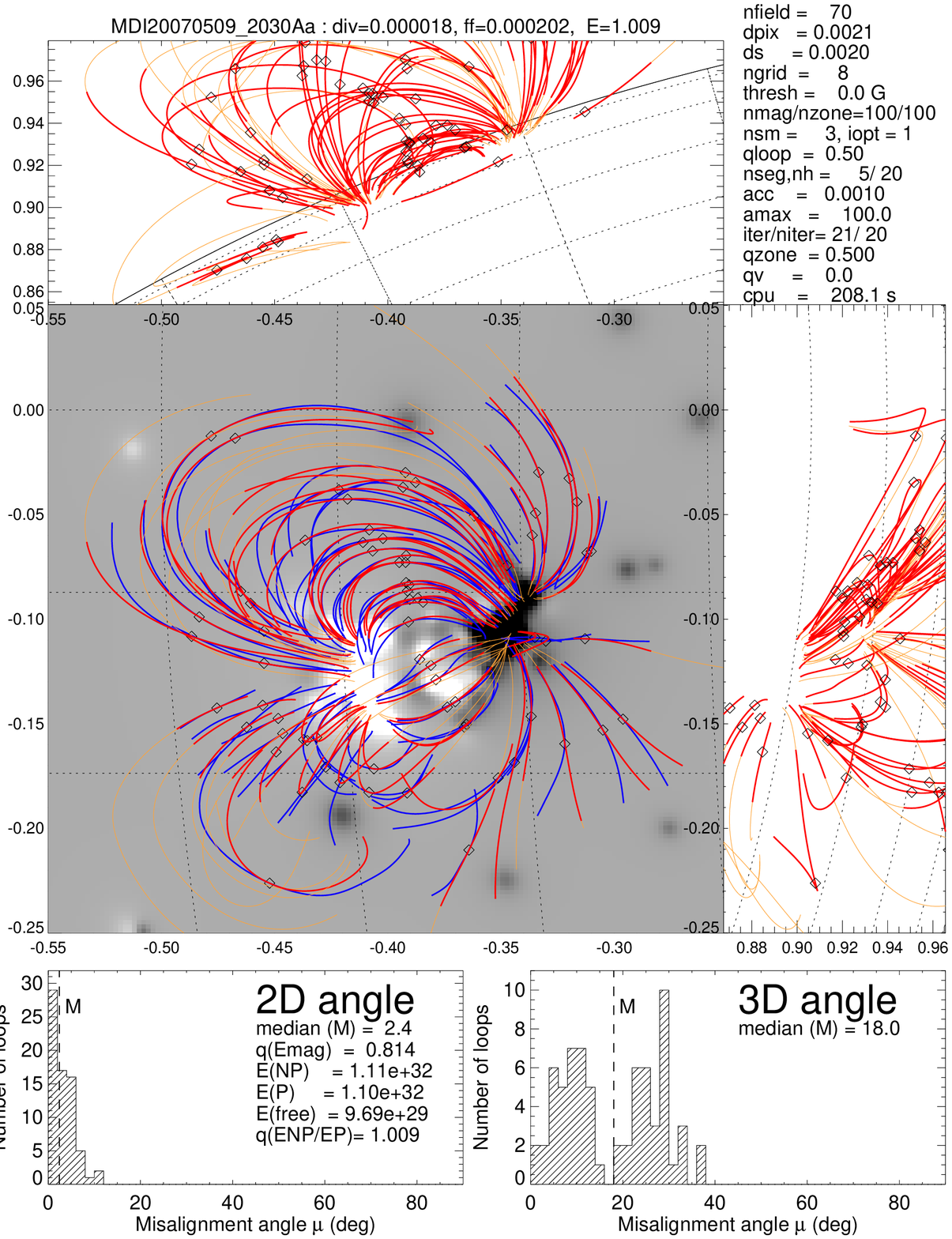}
\caption{Forward-fitting of the NLFFF model to active region 
B (2007 May 9), using STEREO data, in similar representation as Figure 4.}
\end{figure}

\begin{figure}
\plotone{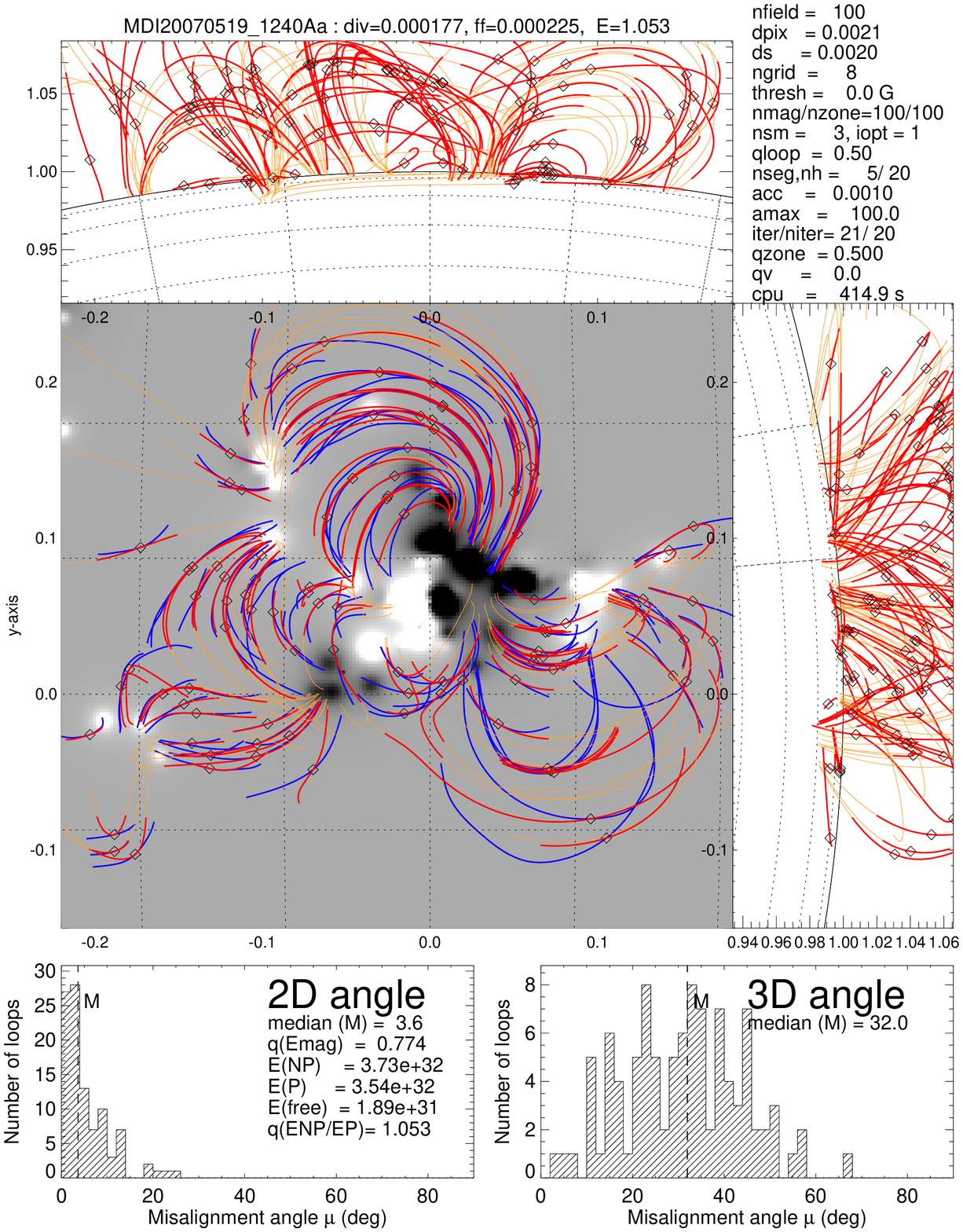}
\caption{Forward-fitting of the NLFFF model to active region 
C (2007 May 19), using STEREO data, in similar representation as Figure 4.}
\end{figure}

\begin{figure}
\plotone{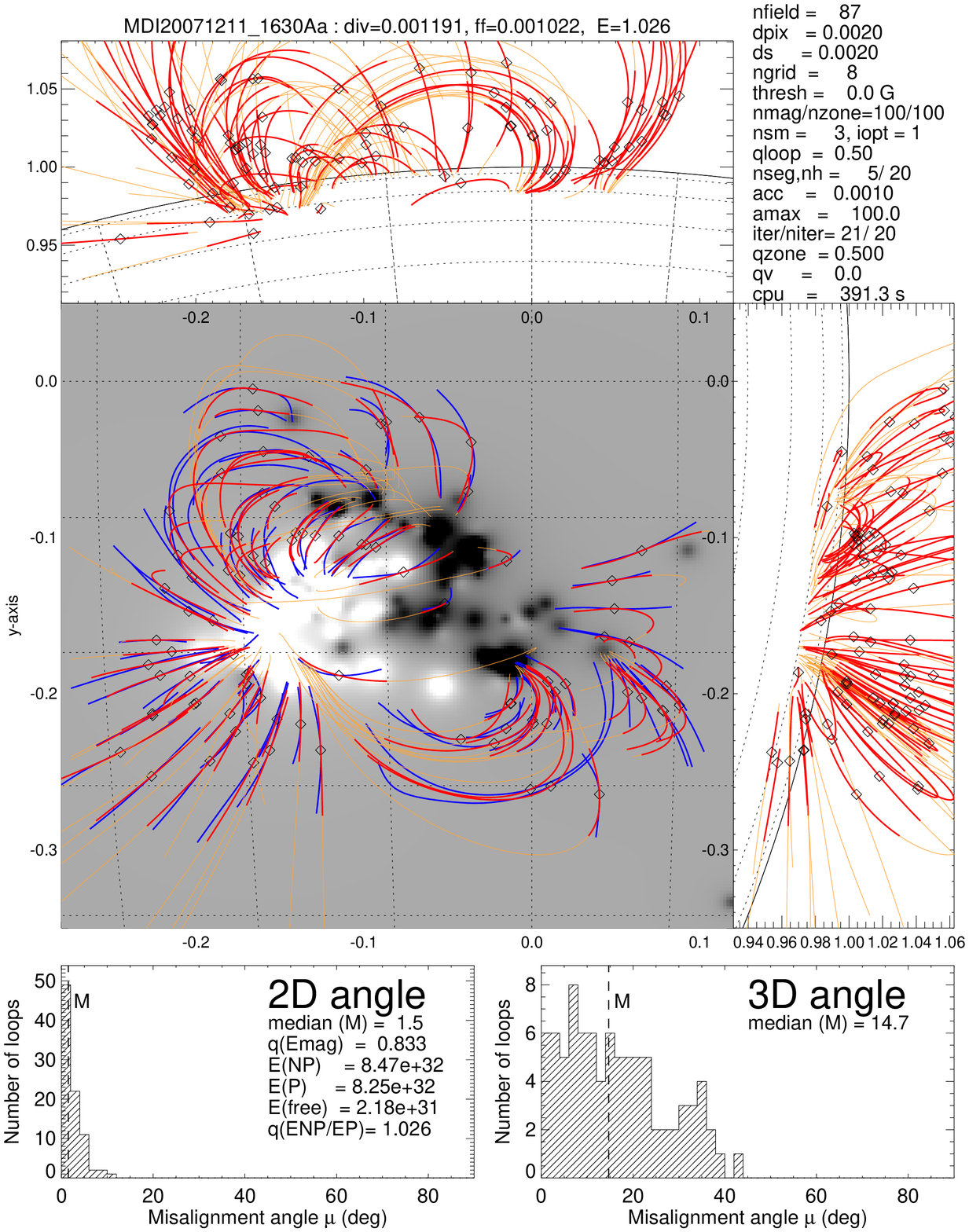}
\caption{Forward-fitting of the NLFFF model to active region 
D (2007 Dec 11), using STEREO data, in similar representation as Figure 4.}
\end{figure}

\begin{figure}
\plotone{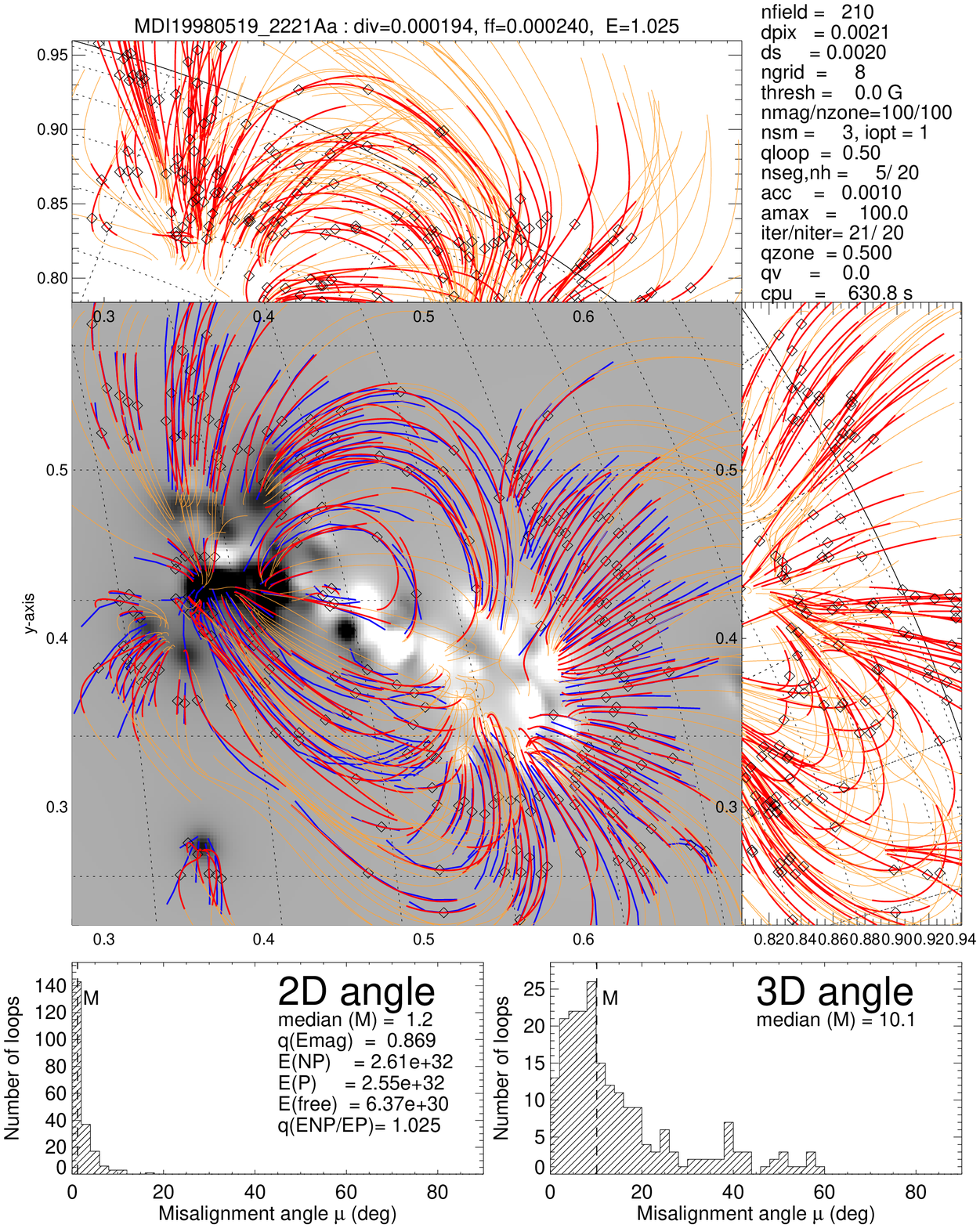}
\caption{Forward-fitting of the NLFFF model to active region 
E (1998 May 19), using manually traced loops from TRACE data,
in similar representation as Figure 4.}
\end{figure}

\begin{figure}
\plotone{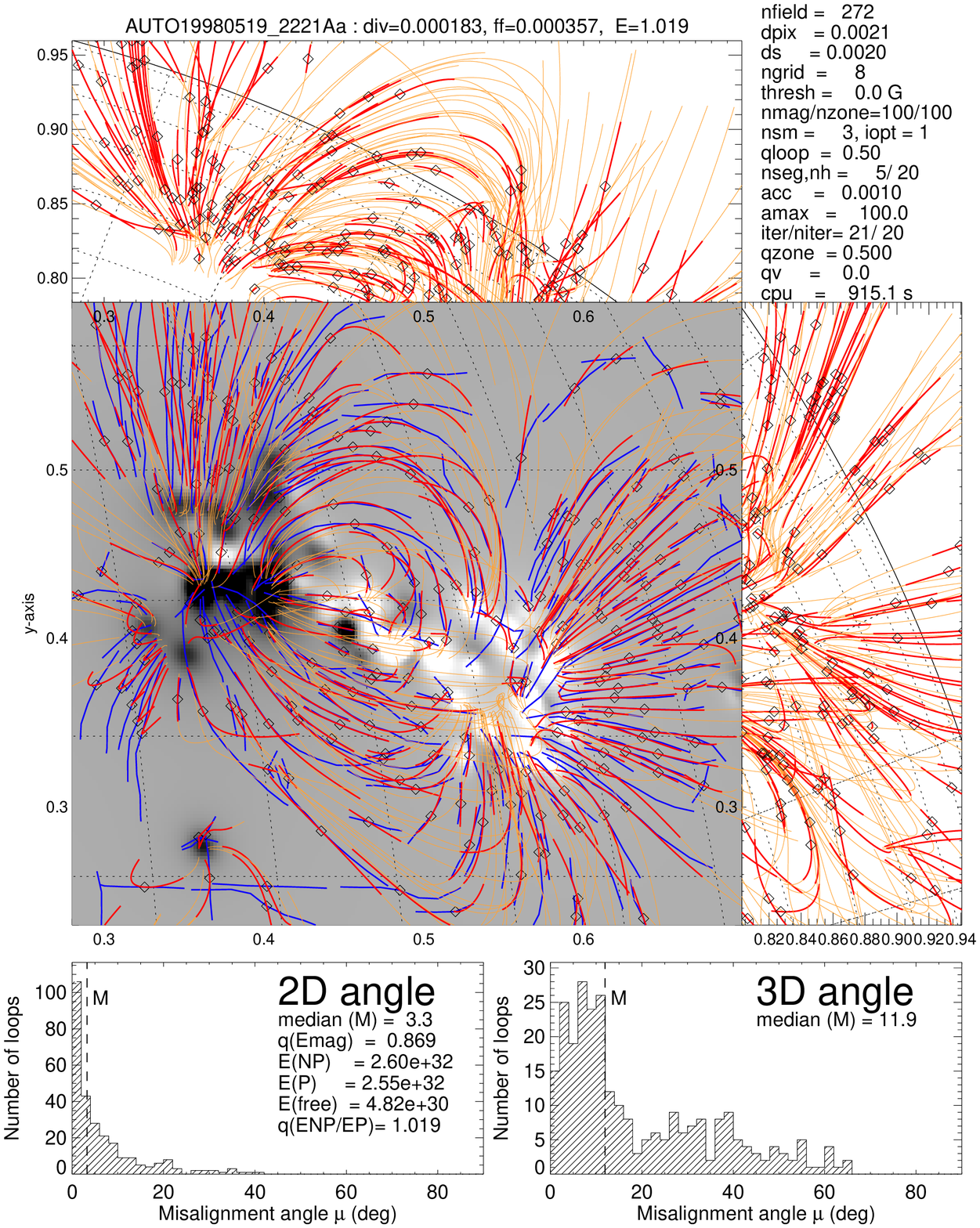}
\caption{Forward-fitting of the NLFFF model to active region 
F (1998 May 19), using automatically traced loops from TRACE data,
in similar representation as Figure 4.}
\end{figure}

\begin{figure}
\plotone{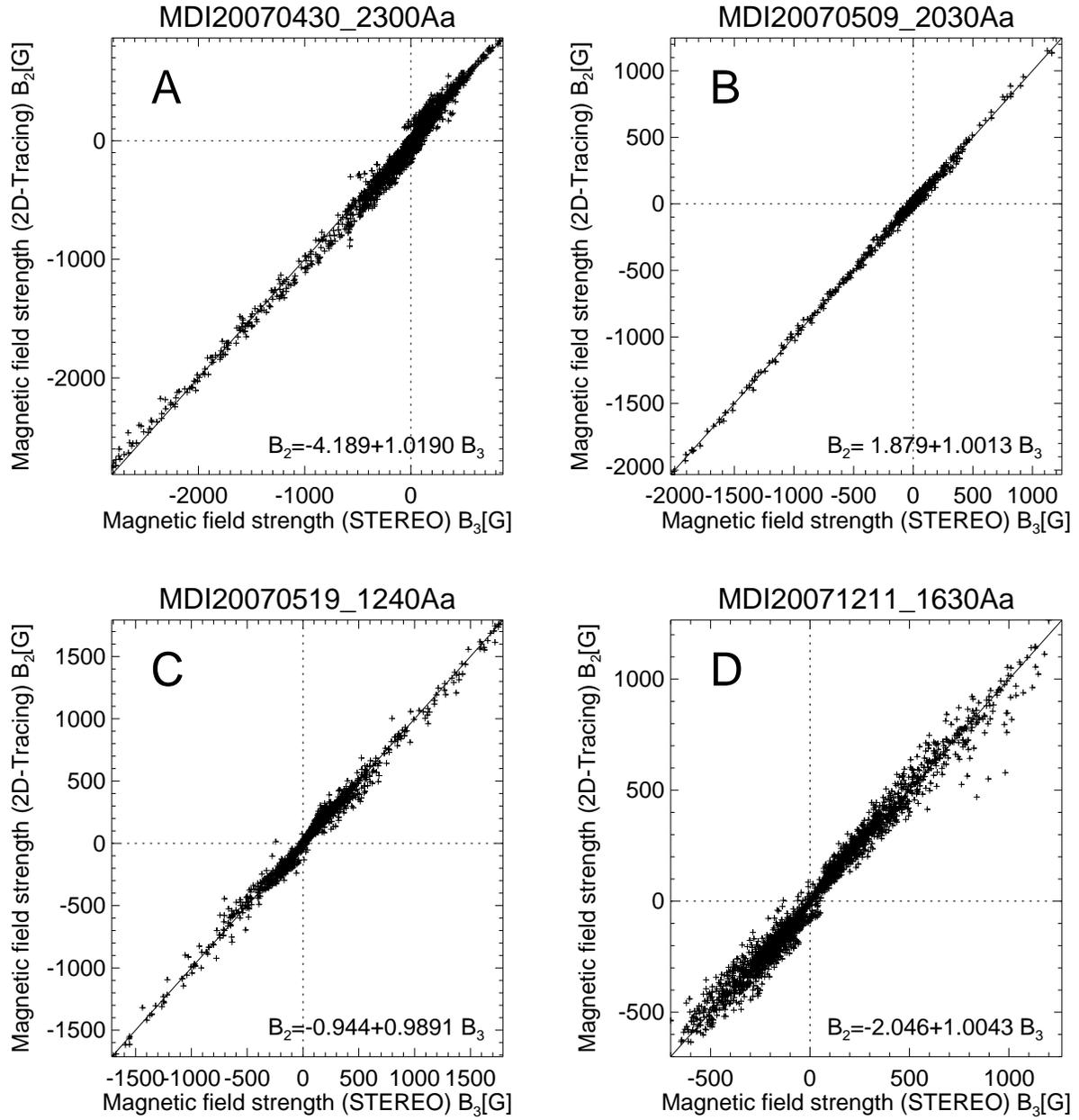}
\caption{Magnetic field strength $B(x,y,z=1)$ in a plane
tangential to solar surface (at $z=1.0$ solar radii from Sun center),
calculated from NLFFF fitting to stereoscopic data (x-axis) versus 
fitting to 2D loop tracing data (y-axis). Each scatterplot contains
the datapoints from a field-of-view that encompasses the full EW 
extent and the middle 10\% of the NS extent shown in Figures 4 to 9.}
\end{figure}

\begin{figure}
\plotone{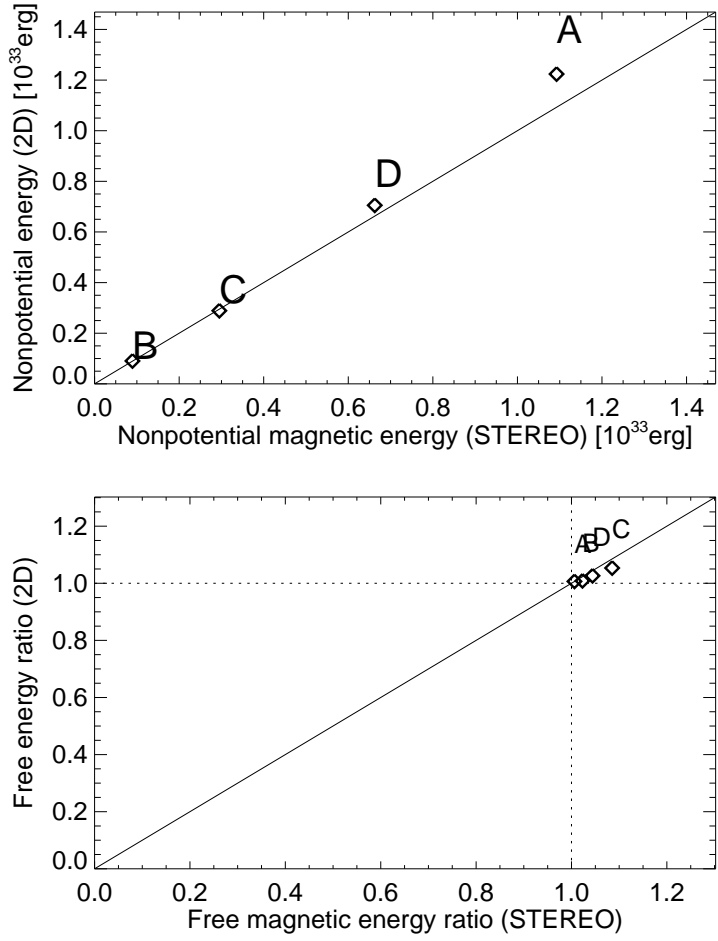}
\caption{Nonpotential magnetic energy (top panel) and ratio of nonpotential
to potential energy (bottom) integrated over each active region (A, B, C, D)
over a height range of $\Delta h=0.015$ solar radii above the photosphere 
and a field-of-view as shown in Figures 4 to 9.}
\end{figure}

\end{document}